\documentclass[aps,prl,twocolumn,superscriptaddress,groupedaddress]{revtex4}  
\usepackage{graphicx}  
\usepackage{dcolumn}   
\usepackage{bm}        
\usepackage{amssymb}   

\begin{document}



\title{Optically addressing single rare-earth ions in a nanophotonic cavity}

\author{Tian Zhong}
\email{tzh@uchicago.edu}
\affiliation{Kavli Nanoscience Institute and Thomas J. Watson, Sr., Laboratory of Applied Physics, California Institute of Technology, Pasadena, California 91125, USA.}
\affiliation{Institute for Quantum Information and Matter, California Institute of Technology, Pasadena, California 91125, USA.}
\affiliation{Institute for Molecular Engineering, University of Chicago, Chicago, IL 60637, USA.}

\author{Jonathan M. Kindem}
\affiliation{Kavli Nanoscience Institute and Thomas J. Watson, Sr., Laboratory of Applied Physics, California Institute of Technology, Pasadena, California 91125, USA.}
\affiliation{Institute for Quantum Information and Matter, California Institute of Technology, Pasadena, California 91125, USA.}

\author{John G. Bartholomew}
\affiliation{Kavli Nanoscience Institute and Thomas J. Watson, Sr., Laboratory of Applied Physics, California Institute of Technology, Pasadena, California 91125, USA.}
\affiliation{Institute for Quantum Information and Matter, California Institute of Technology, Pasadena, California 91125, USA.}

\author{Jake Rochman}
\affiliation{Kavli Nanoscience Institute and Thomas J. Watson, Sr., Laboratory of Applied Physics, California Institute of Technology, Pasadena, California 91125, USA.}
\affiliation{Institute for Quantum Information and Matter, California Institute of Technology, Pasadena, California 91125, USA.}

\author{Ioana Craiciu}
\affiliation{Kavli Nanoscience Institute and Thomas J. Watson, Sr., Laboratory of Applied Physics, California Institute of Technology, Pasadena, California 91125, USA.}
\affiliation{Institute for Quantum Information and Matter, California Institute of Technology, Pasadena, California 91125, USA.}

\author{Varun Verma}
\affiliation{National Institute of Standards and Technology, 325 Broadway, MC 815.04, Boulder, Colorado 80305, USA}

\author{Sae Woo Nam}
\affiliation{National Institute of Standards and Technology, 325 Broadway, MC 815.04, Boulder, Colorado 80305, USA}

\author{Francesco Marsili}
\affiliation{Jet Propulsion Laboratory, California Institute of Technology, 4800 Oak Grove Drive, Pasadena, California 91109, USA}

\author{Matthew D. Shaw}
\affiliation{Jet Propulsion Laboratory, California Institute of Technology, 4800 Oak Grove Drive, Pasadena, California 91109, USA}

\author{Andrew D. Beyer}
\affiliation{Jet Propulsion Laboratory, California Institute of Technology, 4800 Oak Grove Drive, Pasadena, California 91109, USA}

\author{Andrei Faraon}
\email{faraon@caltech.edu}
\affiliation{Kavli Nanoscience Institute and Thomas J. Watson, Sr., Laboratory of Applied Physics, California Institute of Technology, Pasadena, California 91125, USA.}
\affiliation{Institute for Quantum Information and Matter, California Institute of Technology, Pasadena, California 91125, USA.}
\date{\today}

\begin{abstract}
We demonstrate optical probing of spectrally resolved single Nd$^{3+}$ rare-earth ions in yttrium orthovanadate (YVO$_4$). The ions are coupled to a photonic crystal resonator and show strong enhancement of the optical emission rate via the Purcell effect resulting in near-radiatively-limited single photon emission. The measured high coupling cooperativity between a single photon and the ion allows for the observation of coherent optical Rabi oscillations. This could enable optically controlled spin qubits, quantum logic gates, and spin-photon interfaces for future quantum networks.
\end{abstract}


\maketitle

Rare-earth dopants in solids exhibit long-lived coherence in both the optical and spin degrees of freedom \cite{Thiel, Sun}. The effective shielding of 4f electrons leads to optical and radio-frequency transitions with less sensitivity to noise in their crystalline surroundings at cryogenic temperatures. Significant progress in rare-earth based quantum technologies has led to ensemble-based optical quantum memories \cite{Tittel, Riedmatten, Hedges, Zhongsci} and coherent transducers \cite{Williamson}, with promising performance as quantum light-matter interfaces for quantum networks. On the other hand, addressing single ions has remained an outstanding challenge, with the progress hindered by the long optical lifetimes of rare-earth ions and resultant faint photoluminescence (PL). So far, only a few experiments have succeeded in isolating individual praseodymium \cite{KolesovNC, Utikal, Nakamura}, cerium \cite{Kolesov, Siyushev, Xia}, and erbium \cite{Yin, Dibos} ions, though the majority of them did not probe ions via their 4f-4f optical transitions. Recently, several works have demonstrated significant enhancement of spontaneous emissions of rare-earth emitters coupled to a nanophotonic cavity \cite{Zhong, Zhongcavprot, Zhongsci, Dibos}, among which \cite{Zhong, Zhongsci} also showed negligible detrimental effect on the coherence properties of ions in nanodevices. These results point at a viable approach to efficiently detect and coherently control individual ions in a chip-scale architecture. 

Here we demonstrate a nanophotonic platform based on a yttrium orthovanadate (YVO$_4$) photonic crystal nanobeam resonator coupled to spectrally resolved individual neodymium (Nd$^{3+}$) ions. While the system acts as an ensemble quantum memory when operating at the center of the inhomogeneous line \cite{Zhongsci}, it also enables direct optical addressing of single Nd$^{3+}$ in the tails of the inhomogeneous distribution, which show strongly enhanced, near-radiatively-limited single photon emissions. A measured vacuum Rabi frequency of 2$\pi\times$28.5 MHz significantly exceeds the linewidth of a Nd$^{3+}$ ion, allowing for coherent manipulation of spins with optical pulses. Unlike prior experiments \cite{KolesovNC, Utikal, Nakamura, Kolesov, Siyushev, Xia}, this technique does not hinge on the spectroscopic details of a specific type of ion and can be readily extended to other rare-earths or defect centers. The technique opens up new opportunities for spectroscopy on single ions that are distinct from conventional ensemble measurements, which enables probes for the local nanoscopic environment around individual ions and may lead to new quantum information processing, interconnect and sensing devices.

\begin{figure}[ht]
\includegraphics[width=0.45\textwidth]{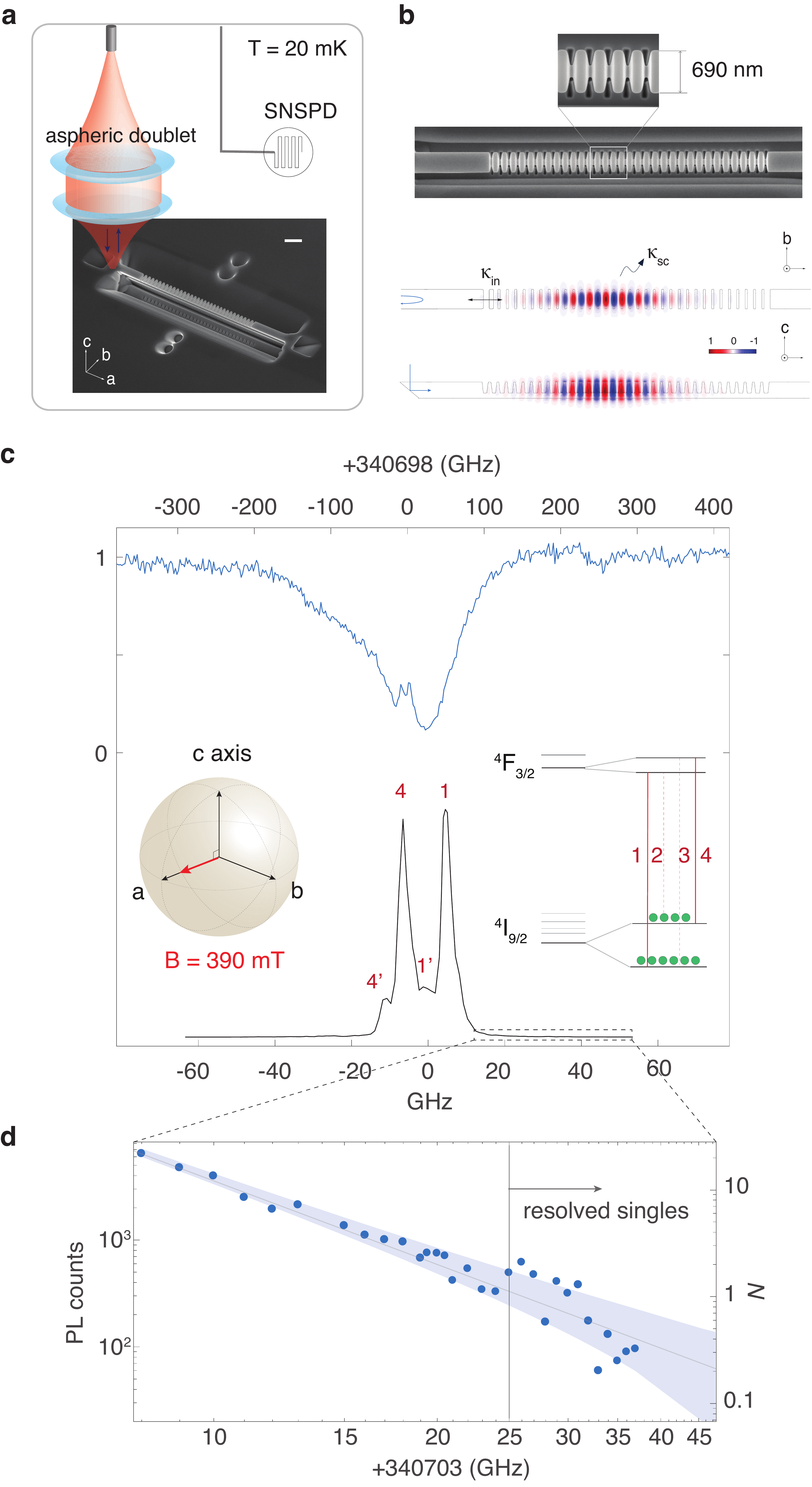}
\caption{\label{fig:epsart} (a) Schematics of the experiment in a dilution refrigerator. Scale bar is 1$\mu$m. (b) SEM images of the one-sided nanobeam photonic crystal cavity in YVO$_4$ fabricated using FIB. Lower panel shows the simulated the TM fundamental mode profile, which has the polarization aligned to the dipoles of Nd$^{3+}$ along the crystallographic c-axis. (c) Cavity reflection spectrum (upper) and Nd$^{3+}$ photoluminescence (PL) spectrum (lower). Insets show the applied magnetic field and resulting Zeeman levels and transitions. PL from ions in the bulk substrate (1' and 4') appear red-shifted from ions coupled to the cavity (1 and 4). (d) Atomic spectra density versus detuning on the shorter wavelength tail of the inhomogeneous distribution. The shaded area shows the projected atomic shot noise.}
\end{figure}

Our experiment builds upon a triangular nanobeam photonic crystal resonator ~\cite{Zhong,Zhongfab} that was fabricated in a nominally 50 parts per million (ppm) doped Nd$^{3+}$:YVO$_4$ crystal using focused ion beam (FIB) milling \cite{Zhongfab}. The device is a one-sided cavity, as the input (left mirror in Fig.1(a,b)) has a lower reflectivity. The optical coupling in/out of the device was implemented via a 45$^{\circ}$-angled coupler \cite{Zhong}. An aspheric doublet mode matches the single mode fiber to the nanobeam waveguide (Fig.1(a)). The coupling efficiency was optimized to 19\% (from fiber to waveguide) using a 3-axis nano-positioner. The nanocavity fundamental mode volume is $V_{\rm mode}$ = 0.056 $\mu$m$^3$ (simulated) with a measured quality factor $Q$ = 3,900 (energy decay rate $\kappa=2\pi\times$90 GHz). The waveguide-cavity coupling $\kappa_{\rm in}$ through the input mirror was 45\% of $\kappa$. The device was cooled to $\sim$20 mK base temperature in a dilution refrigerator, though the actual ensemble temperature was estimated to be around 500 mK (by comparing the ground Zeeman level populations from the PL spectra). The elevated temperature was attributed to the very small thermal conductance in the nanobeam. This limitation has already manifested in previous sub-Kelvin bulk sample measurements \cite{Kukharchyk}, and was even more demanding for measuring nanodevices in the current case. The laser for probing the ions was modulated by two double-pass acousto-optic (AOM) modulators, and delivered to the sample via a single-mode fiber. The reflected signal from the device was sent via a circulator to a superconducting nanowire single photon detector (SNSPD) that measured a 82\% detection efficiency at 880 nm and $<$2 Hz dark counts ~\cite{Zhongsci}. The SNSPD was mounted in the same fridge at the 100 mK stage. The overall photon detection efficiency including transmission from the cavity to the detector and the detector efficiency was 3.6\% (Supplemental material \cite{SM}).

A typical cavity reflection spectrum when tuned nearly on resonance with the Nd$^{3+}$ $^4F_{3/2} (Y_1)-^4I_{9/2}(Z_1)$ transition at 880 nm is shown in Fig.1(c). A 390 mT magnetic field was applied along the crystallographic a-axis of YVO$_4$, giving rise to split Zeeman levels and four possible optical transitions \cite{Hastings} (labelled 1-4) shown in the inset. Symmetry considerations impose that the 2, 3 cross transitions are forbidden, and the 1, 4 transitions are close to cyclic \cite{AfzeliusYVO, Zhongsci}. The PL spectrum (with a 200-ns pulsed resonant excitation) is shown in the lower part of Fig.1(c). Two weak lines labelled 1' and 4' were identified as emissions from Nd$^{3+}$ ions in the bulk substrate, which are red-detuned from ions coupled to the cavity by 2.5 GHz. This shift is due to a static strain in the nanobeam, which makes it easier to spectrally separate the ions in the cavity from the bulk. For subsequent experiments, we focus on the shorter wavelength tail of the inhomogeneous distribution. Figure 1(d) plots the resonant PL against detuning from the peak of line 1 (340703.0 GHz). The PL and thus the atomic spectral density ($N$ ions per excitation pulse bandwidth) fits with a power law of $N\propto\Delta^{-2.9}$, where $\Delta$ is the detuning. The 2.9 power exponent indicates an inhomogeneous broadening mechanism due to strain by dislocation according to the asymptotic form in \cite{Stoneham}. Statistical fine structures (SFS) \cite{Moerner} were also evident. By fitting the SFS with the projected shot noise of $N$ ($\sqrt{N}$ indicated as the shaded area), it is projected that discrete single ion spectra ($N<1$) emerge at a detuning $>$25 GHz. 

\begin{figure*}[ht]
\includegraphics[width=0.95\textwidth]{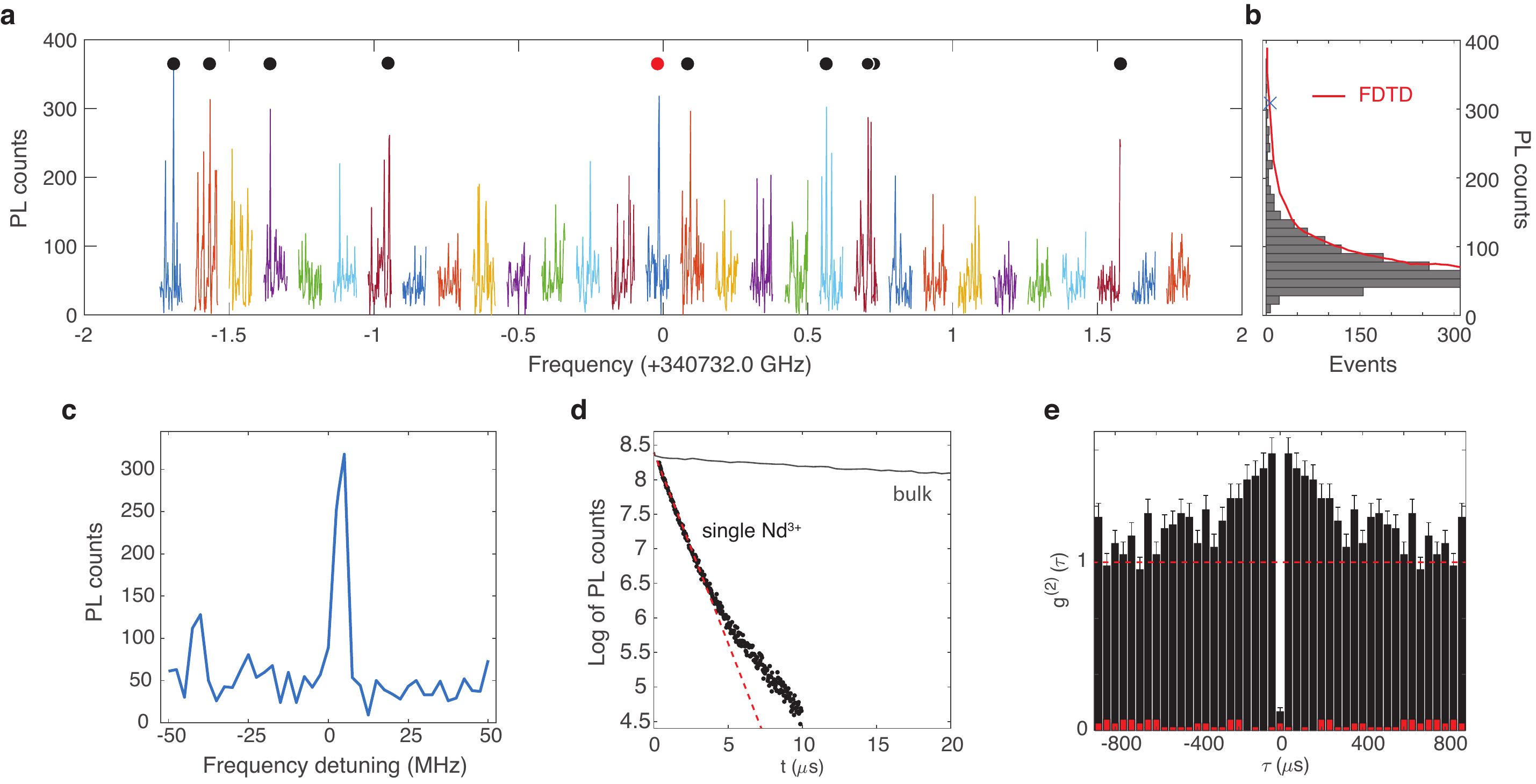}
\caption{\label{fig:wide} (a) Photoluminescence (PL) spectrum swept over 3 GHz around $\Delta\sim$30 GHz. Isolated peaks marked by filled circles correspond to individual Nd$^{3+}$ ions. Each color represents a different laser scan. (b) Histogram of PL intensities from the ensemble of Nd$^{3+}$. The red curve is a FDTD simulation of the expected distribution given the different position of each ion inside the cavity. (c) PL spectrum of the ion labelled with the red circle in Fig.2(a). (d) PL decay of the  ion (black) with a fitted T$_1$ = 2.1$\pm$0.2 $\mu$s (red) compared to T$_1$ = 90 $\mu$s in bulk (grey). (e) Intensity autocorrelation measurement on the single Nd$^{3+}$ showing anti-bunching (g$^2$(0)=0.09$\pm$0.013). The background signal with an off-resonant excitation is in red.}
\end{figure*}

To search for singles, we scanned the frequency of a 200-ns resonant excitation pulse around $\sim$30 GHz blue detuning from the peak of line 1, and measured the PL integrated over 5 $\mu$s after the excitation. The repetition rate of the excitation pulses was 25 kHz, and the integration time was 20 seconds at each frequency. The laser was frequency stabilized to a vacuum-chamber reference cavity attaining a narrowed linewidth of $<$5 kHz and a long-term drift $<$100 kHz/day. Figure 2(a) shows the measured PL over a few GHz range. A handful of peaks, such as the one with close-up in Fig.2(c), were possible single Nd$^{3+}$ ions. The PL intensities were histogrammed in Fig.2(b) to reveal a distribution of ion-cavity coupling strengths, which is in good agreement with that from the finite difference time domain (FDTD) simulation (red). Thus, the PL intensity serves to correlate the coupling strength of each ion with its spatial position relative to the cavity anti-nodes: an ion located at the antinode would have the strongest coupling and show the highest PL. The linewidth of the peak in Fig.2(c) was broadened by the excitation pulse. The actual linewidth of single ions was expected to be considerably narrower. With the laser tuned on-resonance with one of the peaks (marked with a red dot in Fig.2(a)), the intensity autocorrelation measurement using a single detector yielded a g$^{(2)}$(0) = 0.09 $\pm$ 0.013 (Fig.2(e)) with $\sim$0.02 photons generated per pulse, which was normalized to g$^{(2)}$($t$) at large $\tau$. The bunching behavior at $|\tau|<$600$\mu$s was expected from a multilevel emitter \cite{Kurtsiefer, Wu}. The imperfect anti-bunching was partly due to a continuum of ions that is weakly coupled to the cavity, resulting in a background in Fig.2(e). This background was measured with the excitation laser far detuned from the single ion resonance. The optical T$_1$ of this ion was 2.1$\pm$0.2 $\mu$s (Fig.~2(d)), which is strongly enhanced compared to the bulk T$_1$ of 90 $\mu$s. The lifetime enhancement corresponded to a Purcell factor of 111 of the probed Y$_1$-Z$_1$ transition considering a branching ratio of $\beta$=0.38 (the ground state splits into five Kramers doublets Z$_1$-Z$_5$) \cite{SM}. The theoretically maximum Purcell factor was $F\approx\frac{3}{4\pi^2\chi_L^2}(\frac{\lambda}{n_{\rm YVO_4}})^3\frac{Q}{V}=$189 \cite{Purcell, Mcauslan} assuming a perfect alignment of the dipole with the cavity mode and $\chi_L=3n_{\rm YVO_4}^2/(2n_{\rm YVO_4}^2+1)$ is the local correction to the electric field since the ion is less polarizable than the bulk medium \cite{Dung}, where we have used the real cavity approximation (supplemental material \cite{SM}). The discrepancy is attributed to the non-optimal position of the ion, and the actual cavity mode volume being different from simulation because of fabrication imperfections. 

\begin{figure}[h]
\includegraphics[width=0.45\textwidth]{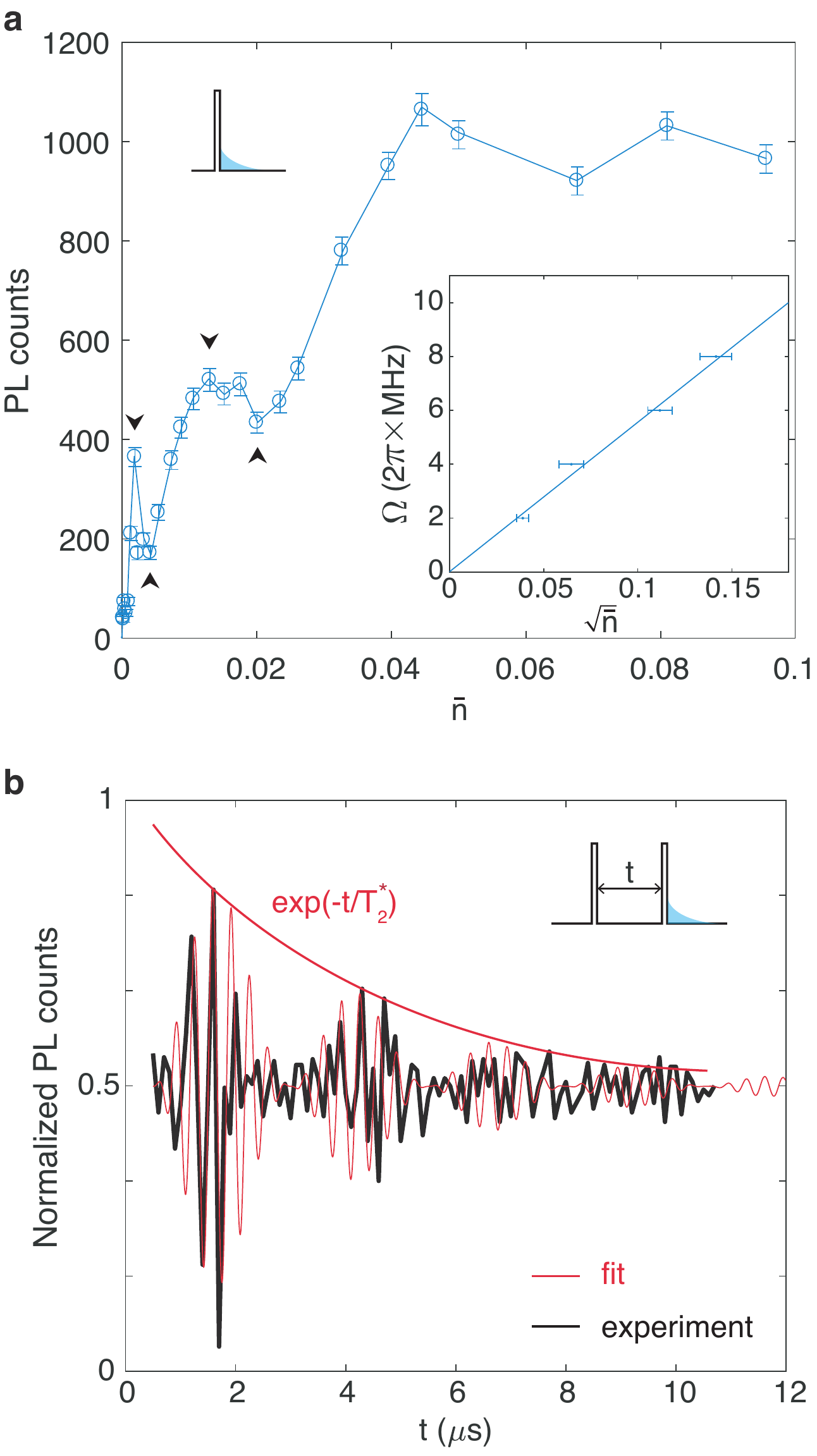}
\caption{\label{fig3} (a) Rabi oscillations of PL following a pulsed resonant excitation with increasing photon number $\bar{n}$. Black arrows point to pulse areas that are integer multiples of $\pi$. The inset plots extracted Rabi frequencies against $\sqrt{\bar{n}}$ with a linear fit. (b) Normalized Ramsey interference fringes. The beating at a frequency of 740 kHz is consistent with the superhyperfine coupling between the Nd$^{3+}$ ion and surrounding Y.}
\end{figure}
The small mode volume of the nanocavity results in a significant enhancement of the coupling strength $g_0$. Focusing on the ion in Fig.2(c), Fig.3(a) plots the PL excited by a square 250-ns resonant pulse with increasing cavity mean photon number $\bar{n}$. The value of $\bar{n}$ was calculated from the input pulse energy, all losses in the setup up to the device, and coupling rates of the photonic crystal mirrors (Supplemental material \cite{SM}). The PL shows Rabi oscillations similar to an optical nutation \cite{Gerhardt}. The inset plots the extracted Rabi frequencies $\Omega$ versus square root of $\bar{n}$ from the peaks (corresponding to odd integer of $\pi$ pulse areas) and valleys (even integer of $\pi$ pulses) of the Rabi oscillations. The fitted slope corresponds to $g_0=\Omega/2\sqrt{\bar{n}}$ = 2$\pi\times$28.5$\pm$5.2 MHz. The theoretical maximum $g_0$ is $\mu/n_{\rm YVO_4}\sqrt{\omega_0/2\hbar\epsilon_0V}=2\pi\times$52.7 MHz \cite{Mcauslan}, where $\mu$=1.59$\times$10$^{-31}$ C$\cdot$m is the transition dipole moment (supplemental material \cite{SM}), $n_{\rm YVO_4}$=2.1785 is the refractive index of YVO$_4$, $\omega_0$ is the transition frequency, and $\epsilon_0$ is the vacuum permittivity. The measured $g_0$ is orders of magnitude stronger than the linewidth of the emitter, which makes possible the use of hard optical pulses \cite{Pryde} to coherently control each single ion. Next, we applied two $\pi/2$ pulses to measure the Ramsey interference as shown in Fig.3(b). The normalized Ramsey fringes (subtracting a T$_1$ decay background) reveal a clear beating, which most likely corresponds to the superhyperfine interactions between Nd electronic spins and the nuclear spins of yttrium in YVO$_4$ \cite{SM}. The measured superhyperfine beating, confirmed by the two-pulse photon echo measurement (Supplemental material \cite{SM}), was 740 kHz, which is consistent with the calculations based on the gyromagnetic ratio of yttrium (Y) nuclear spins of $\sim$2.1 MHz/T \cite{Car, SM}. At a relatively strong field of 390 mT, the Nd-Y superhyperfine structure is dominated by the yttrium nuclear magnetic moment (Supplemental material \cite{SM}), as also observed in Nd$^{3+}$:Y$_2$SiO$_5$ \cite{Usmani}. The decay of the Ramsey fringe envelope can be fitted empirically to extract a T$_2^*$ = 4.0$\pm$0.2 $\mu$s. From that, the spectral indistinguishability is calculated as $T_2/(2T_1)$=0.952, indicating that the linewidth of this ion approaches the radiatively-limited regime. 

\begin{figure}[ht]
\includegraphics[width=0.45\textwidth]{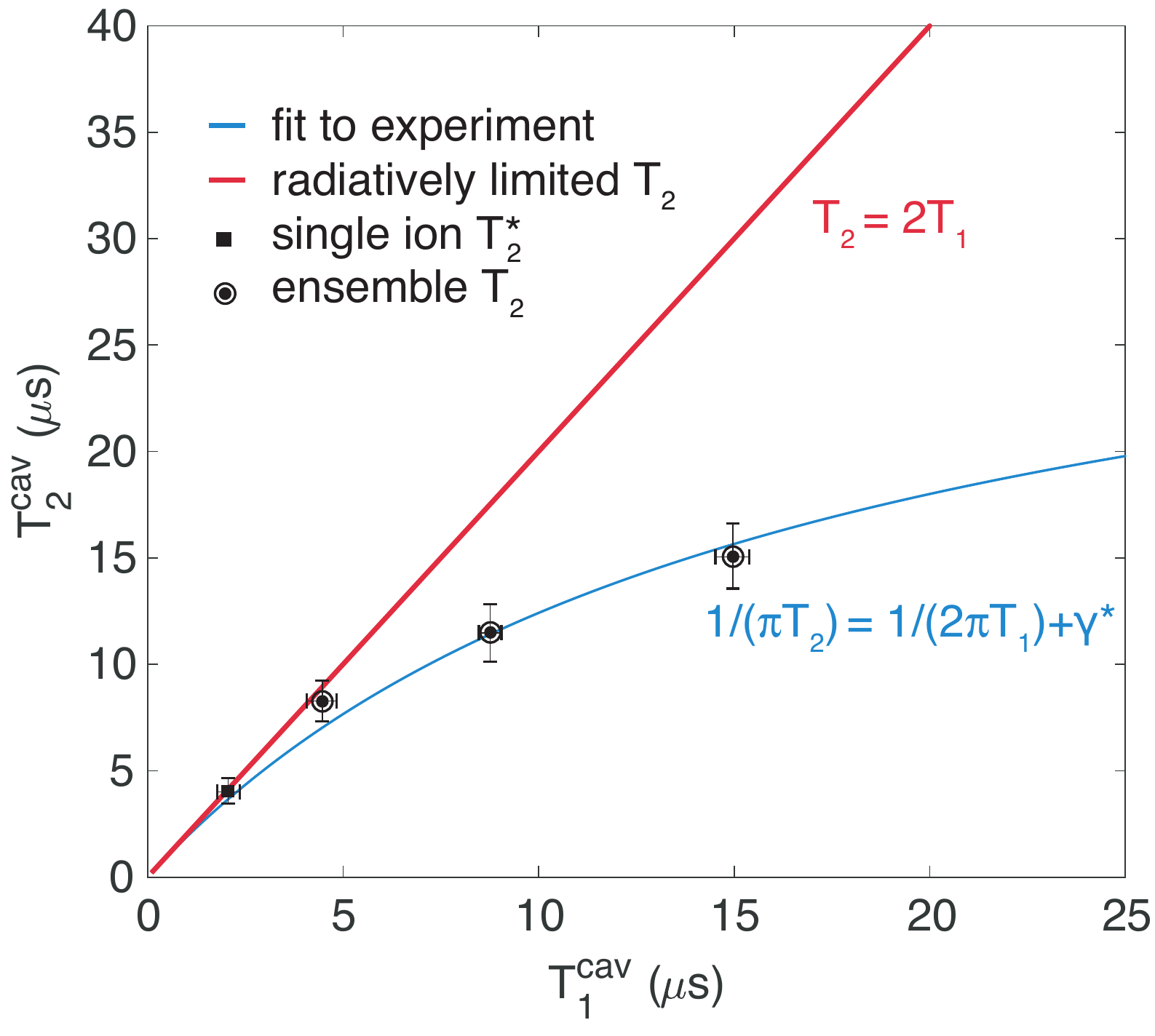}
\caption{\label{fig4} Measured and theoretical optical coherences for Nd$^{3+}$ coupled to the cavity with varying detuning and Purcell enhancement. The red line is the radiatively-limited T$_2$ time. The single Nd$^{3+}$ on resonance with the cavity (square) exhibits a near radiatively-limited linewidth with a spectral indistinguishability $>95\%$.}
\end{figure}

The use of single rare-earth ions as spin-photon interfaces to entangle remote quantum nodes requires each emitter's linewidth to be radiatively limited. To further characterize the coherence of the ions coupled to the cavity, we performed additional ensemble two-pulse photon echo measurements when the emitters have different detunings from cavity resonance. The ensemble T$_2$ times are plotted against optical T$_1$ in Fig.4, including the single ion T$_2^*$ data denoted by a square. The experimental data were fitted with the relationship $1/(\pi T_2)=1/(2\pi T_1)+\gamma^*$, where $\gamma^*$ is the pure dephasing rate. The fit (blue curve) gives a $\gamma^*$=9.7$\pm$0.6 kHz. While slow, this dephasing rate was attributed to the superhyperfine interactions since it closely matches the superhyperfine-limited T$_2$ in Nd$^{3+}$:YVO$_4$ \cite{Sun}. The contribution from Nd$^{3+}$ spin flip-flops are expected to be small, because the measured T$_2$ in an nominally undoped YVO$_4$ crystal (residue doping estimated at 0.2 parts per million) was comparable to that measured in the current device (Supplemental material \cite{SM}). 

The full radiatively limited T$_2$=2T$_1$ is plotted in red. With weak enhancement when the ions are detuned from the cavity, the ions exhibit poor indistinguishabilities as indicated by the sizeable gap between the red and blue curves. Only when the ions are coupled to the cavity resonantly, do they become radiatively limited.  A similar approach has been used to improve the single photon indistinguishabilities of quantum dots \cite{Grange}. To increase the indistinguishability, improving the cavity quality factor to further reduce T$_1$ would be a straightforward step, which would also allow the device to operate at higher temperatures with stronger dephasing while still achieving radiatively limited emissions. The current linewidth of the single emitter was based on T$_2^*$ values measured over a few $\mu$s (Fig.3(b)). For longer time scales (100 $\mu$s to ms), reducing the slow optical spectral diffusion could help to maintain a high indistinguishability, as desired by quantum memories for long-distance quantum network. In that regard, using rare-earth emitters in hosts with weaker nuclear spin baths or non-Kramers ions with weaker superhyperfine couplings and operating at a zero-first-order-Zeeman (ZEFOZ) point \cite{Fraval04} may offer some advantages.

In conclusion, we have optically detected single Nd$^{3+}$ ions coupled to a nanophotonic cavity, which enhanced the emitter spontaneous emission rate to the extent that the linewidth of the emitter became radiatively-limited. Optical Rabi oscillations of the single Nd$^{3+}$ yielded a $g_0$=2$\pi\times$28.5 MHz, and a linewidth of 12.5 kHz ($\gamma_h$=1/($\pi$T$_2$), where T$_2$=25.4 $\mu$s is the emitter homogeneous linewidth without cavity enhancement (Supplemental material \cite{SM})). Given the cavity decay of $\kappa$=2$\pi\times$90 GHz, the single ion cooperativity is $4g_0^2/\kappa\gamma_h$=2.9. This value could be improved significantly by using cavities with higher Q ($\times$10 higher Q devices already demonstrated in \cite{Zhongfab} would attain an indistinguishability $>$99.5\% and C$\sim$30), thus making feasible the implementation of high-fidelity non-destructive detection of optical photons with a single rare-earth ion \cite{Brien16}. Nevertheless, questions remain regarding the spin coherence and the qubit storage time of single ions \cite{Wolfowicz}, and spectral diffusion occurring at longer time scales. When two spectrally resolved ions are nearby, their dipole-dipole interaction can also be probed \cite{Longdell}. Single rare-earth ions could be used to probe the field and temperature of its nanoscopic surroundings. Finally, the large inhomogeneous linewidth of the emitters may facilitate spectral multiplexing of individual quantum emitters for expanded bandwidth of quantum communication networks.

\begin{acknowledgments}
This work was funded by a National Science Foundation (NSF) Faculty Early Career Development Program (CAREER) award (1454607), the AFOSR Quantum Transduction Multidisciplinary University Research Initiative (FA9550-15-1-002), and the Defense Advanced Research Projects Agency Quiness program (W31P4Q-15-1- 0012). Equipment funding was also provided by the Institute of Quantum Information and Matter, an NSF Physics Frontiers Center with support from the Moore Foundation. The device nanofabrication was performed in the Kavli Nanoscience Institute at the California Institute of Technology. Part of the research was carried out at the Jet Propulsion Laboratory, California Institute of Technology, under a contract with the National Aeronautics and Space Administration. The authors would like to acknowledge Neil Sinclair, Ruffin Evans, Alp Sipahigil, Charles W. Thiel, Jeffrey Thompson for useful discussions.  
\end{acknowledgments}

\clearpage

\newcommand{\beginsupplement}{%
        \setcounter{table}{0}
        \renewcommand{\thetable}{S\arabic{table}}%
        \setcounter{figure}{0}
        \renewcommand{\thefigure}{S\arabic{figure}}%
     }

\onecolumngrid

      \beginsupplement

\section{Supplementary material for optically addressing single rare-earth ions in a nanophotonic cavity}

\section{Details on the experimental setup}
\noindent Figure S1 illustrates the experimental setup with more details. The Nd$^{3+}$:YVO$_4$ sample crystal was soldered with indium onto a copper plate that was mounted on top of a 3-axis nanopositioner, and was thermally connected to the 20 mK base plate of the dilution refrigerator. Cavity tuning was realized by gas condensation using N$_2$ gas. The gas tube (brown line in Fig. S1) was thermally anchored to the 3.8 K stage. When performing gas tuning, a heater on the gas tube heats it up to $>$30 K to allow gas to flow through. The heater was turned off after tuning. This configuration was used to minimize the heat load generated by the tube, allowing the lowest possible temperature at the sample.

\noindent {\bf Fiber-waveguide coupling efficiency}
The coupling to the waveguide was realized by a 45$^{\circ}$-angled cut into the YVO$_4$ substrate \cite{Zhongsci}. This angled cut allows the vertically focused beam (from a fiber) to be total internally reflected into the nanobeam waveguide. The aspherical doublet has an effective focal length of 10.5 mm on the fiber side and 2.9 mm on the sample side (Fig.S1). The coupling was optimized by fine scanning the focal spot on the sample surface using the nanopositioner and looking for the cavity resonance on a spectrometer. The overall fiber-waveguide coupling efficiency was characterized by measuring the reflection of a pulse far off resonance with the cavity (i.e. in the photonic bandgap). The pulse propagated from point 1 (marked in Fig.S1) to 2, 3, and was reflected back to 2, then 4. The transmission efficiency from 1 to 2 (64.1\%), and from 2-4 (51.7\%) were directly measured. The only unknown was the coupling efficiency from 2 to 3. This coupling efficiency could be uniquely determined from the total pulse reflection, and was found to be 19\%.

\noindent {\bf Photon collection efficiency} For each photon emitted by an ion into the cavity, the probability of that photon to be transmitted to the coupling waveguide was $\kappa_{in}/\kappa$=45\%. The photon then propagated from 3 to 2 (19\% waveguide-fiber coupling), and from 2 to 4 (79.9\% transmission through all fiber slices/connectors and 64.7\% transmission through the optical circulator), and was finally detected by the 82\%-efficient superconducting nanowire detector. Thus, the overall collection efficiency for a cavity photon was 0.45$\times$0.19$\times$0.80$\times$0.65$\times$0.82 = 3.6\%. To improve this efficiency, further refinement in fabrication is needed to achieve a highly over-coupled cavity with $\kappa_{in}/\kappa>$80\%, which should be within reach since we have achieved symmetric two-sided cavities of Q as high as 20,000 \cite{Zhongfab}. Furthermore, the use of tapered fiber-waveguide coupling could improve the fiber-to-device coupling to 97\% \cite{Tiecke}. These combined with lower loss fiber components could increase the overall efficiency to $\sim$70\%.
\begin{figure}[ht]
\includegraphics[width=0.45\textwidth]{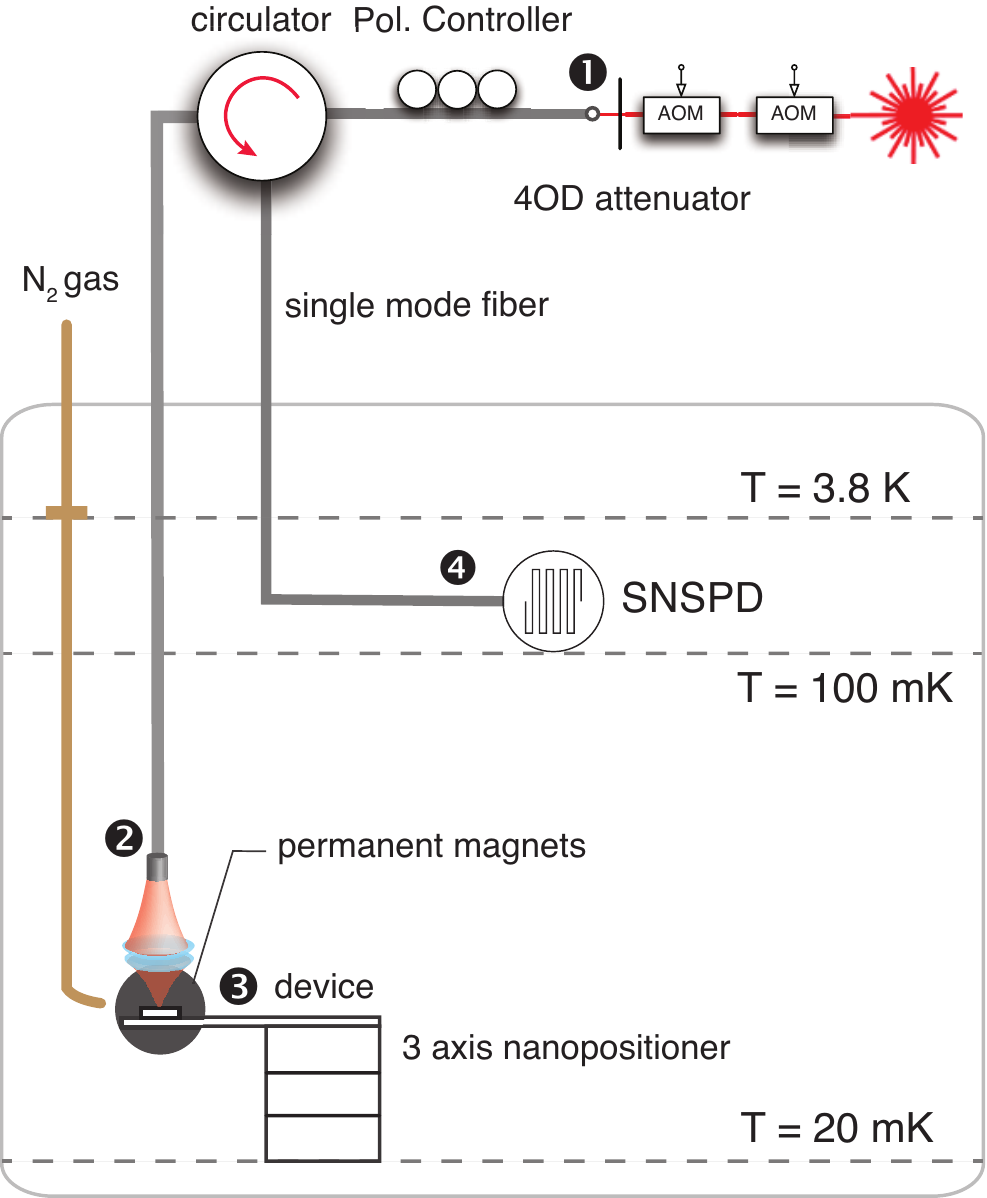}
\caption{Details of the experimental setup. SNSPD: superconducting nanowire single photon detector.}
\end{figure}

\noindent {\bf Cavity mean photon number}
To obtain the cavity mean photon number in Fig.3, we first calculate the peak power of the excitation pulse in the waveguide i.e. P$_{in}$ with knowledge of the transmission from 1 to 2 and the coupling efficiency from 2 to 3. The cavity mean photon number was $\bar{n}=4P_{in}\kappa_{in}/\hbar\omega_0\kappa^2$, where $\kappa_{in} = 2\pi\times$40 GHz was the cavity in-coupling rate, $\kappa=2\pi\times$90 GHz was the total cavity decay rate, and $\omega_0$ is the photon frequency.

\subsection{Optical properties of Nd$^{3+}$ in YVO$_4$ and Purcell factor}
The absorption area $\int \alpha(\nu) d\nu$ for the $^4$I$_{9/2}($Z$_1) \leftrightarrow ^4$F$_{3/2}($Y$_1)$ optical transition in a 10~ppm Nd$^{3+}$:YVO$_4$ sample was measured in zero magnetic field with E $\parallel$ c by Sun \textit{et al.}~\cite{Sun} to be 3.4~cm$^{-1}$~cm$^{-1} = 102$~GHz~cm$^{-1}$. Because the electric dipole transition is heavily $\pi$-polarized, we can calculate the transition oscillator strength from the absorption area $\int \alpha(\nu) d\nu$ using~\cite{DiBartolo1968, Henderson2006}
\begin{eqnarray}
f = 4\pi \epsilon_0 \frac{m_e c}{\pi e^2} \frac{1}{N} \frac{n}{{\chi_{L}}^2} \int \alpha(\nu) d\nu\;\;,
\label{OS}
\end{eqnarray}
where $\epsilon_0$ is the vacuum permittivity, $m_e$ is the mass of the electron, $e$ is the charge on the electron, $c$ is the speed of light, $N$ is the number density, $\chi_L$ is the local field correction, $n$ is the refractive index, and $\alpha(\nu)$ is the absorption coefficient as a function of frequency $\nu$. 

For the 10~ppm Nd$^{3+}$:YVO$_4$ crystal measured by Sun \textit{et al.}~\cite{Sun}, 
$N =1.24\times10^{23}~\textrm{m}^{-3}$ and $n = 2.1785$ for light polarized along the $c$-axis of the crystal. The local field correction factor $\chi_L$ usually takes one of two forms in the literature, depending on whether the virtual cavity or real cavity model is used~\cite{Reid2005, McAuslan2009, Dung2006, Dolgaleva2012}:
\begin{eqnarray}
\chi_L^{(V)} = \frac{n^2+2}{3}\;\;,
\end{eqnarray}
is the virtual cavity model, and 
\begin{eqnarray}
\chi_L^{(R)} = \frac{3n^2}{2n^2+1}\;\;,
\end{eqnarray}
is the real cavity model correction.

Both the virtual and real cavity models are approximations to the full local field correction in that they assume that the field due to the polarization of atoms nearby in the lattice is zero~\cite{Reid2005}. The real cavity model has been shown to be suitable for substitutional ions~\cite{Vries1998} including rare-earth ions in crystalline hosts~\cite{Dolgaleva2012}. In this work, the predictions based on the real cavity model are more consistent with our current knowledge of material and experimental results.

We note that there is inconsistency in the literature regarding oscillator strengths and dipole moments of 4f - 4f transitions for rare-earth ions in crystals because values and expressions are not always explicit as to which local field correction, if any, is assumed. Here we detail our derivations to make it clear the assumptions we have made and how that impacts the theoretical predictions.

When assuming the real cavity model for the local field correction, from Eq.~\ref{OS} we calculate an oscillator strength $f = 3.7\times 10^{-5}$. For the applied field of 390~mT along the a-axis we expect optical transitions 2 and 3 to be forbidden. In this case, transitions 1 and 4 each have an oscillator strength $f = 3.7\times 10^{-5}$.

The radiative lifetime $T_{rad}$ is related to the oscillator strength $f$ by~\cite{DiBartolo1968, Henderson2006}
\begin{eqnarray}
\frac{1}{T_{rad}} = \frac{2 \pi e^2}{\epsilon_0 m_e c} \left(\frac{3n^2}{2n^2+1}\right)^2 \frac{1}{n} \frac{n^2}{\lambda^2} \frac{f}{3} \;\;,
\end{eqnarray}
which gives a value of $T_{rad} = 237~\mu$s. Given the lifetime of the $^4$F$_{3/2}($Y$_1)$ state measured by fluorescence was $T_1 = 90~\mu$s, the branching ratio of emission to the  $^4$I$_{9/2}($Z$_1)$ state is $\beta = 0.38$. 

The transition dipole moment $\mu$ is related to the oscillator strength $f$ by~\cite{DiBartolo1968, Henderson2006}
\begin{eqnarray}
\mu = \sqrt{\frac{\hbar e^2 f}{2 m_e \omega}}\;\;,
\label{mu}
\end{eqnarray}
where $\omega = 2\pi c / \lambda$ is the frequency $^4$I$_{9/2}($Z$_1) \leftrightarrow ^4$F$_{3/2}($Y$_1)$ optical transition. Equation~\ref{mu} differs from the expressions relating $\mu$ to $f$ in ~\cite{Reid2005}. This is because in ~\cite{Reid2005} no field corrections are assumed in relating $f$ to $\int \alpha(\nu) d\nu$. Using Equation~\ref{mu}, we calculate a dipole moment $\mu = 1.59 \times 10^{-31}$~C$\cdot$m.

Given $g_0$ and the cavity energy decay rate $\kappa = 2\pi\times90$~GHz, the lifetime of the Nd$^{3+}$ ion in the cavity is given by
\begin{eqnarray}
T_{cav} = \left(\frac{4g_0^2}{\kappa}+\frac{1-\beta}{T_1}\right)^{-1}\;\;.
\label{Tcav}
\end{eqnarray}
Therefore, the predicted $T_{cav} = 1.25~\mu$s. The Purcell enhancement factor of the resonant transition is derived to be $4g_0^2T_{rad}/\kappa = 189$.

\section{Photon echo measurements}
Two pulse photon echo measurements were performed on an ensemble of ions in the cavity near the center of inhomogneous distribution (e.g. line 1). The cavity resonance was tuned to different frequencies using a gas condensation technique to obtain homogeneous linewidths of ions at varying Purcell enhancement conditions. Fig.S2 plots the photon echo decays at ensemble-cavity detuning of $\delta\sim$22 and $\sim$50 GHz. Oscillations in the echo intensities correspond to syperhyperfine interactions between Nd spins and Y nuclear spins \cite{Usmani} at 740 kHz, which agree with the beat frequency observed in the Ramsey interference fringes. Note that the period of the oscillations appears to be twice long in the Ramsey fringes than in the echo decays, because the photon echo is emitted after twice the delay between two pulses. The T$_2$ were fitted from the linear section of the decay, which started after approximately 4 $\mu$s.

\begin{figure}[h]
\includegraphics[width=0.45\textwidth]{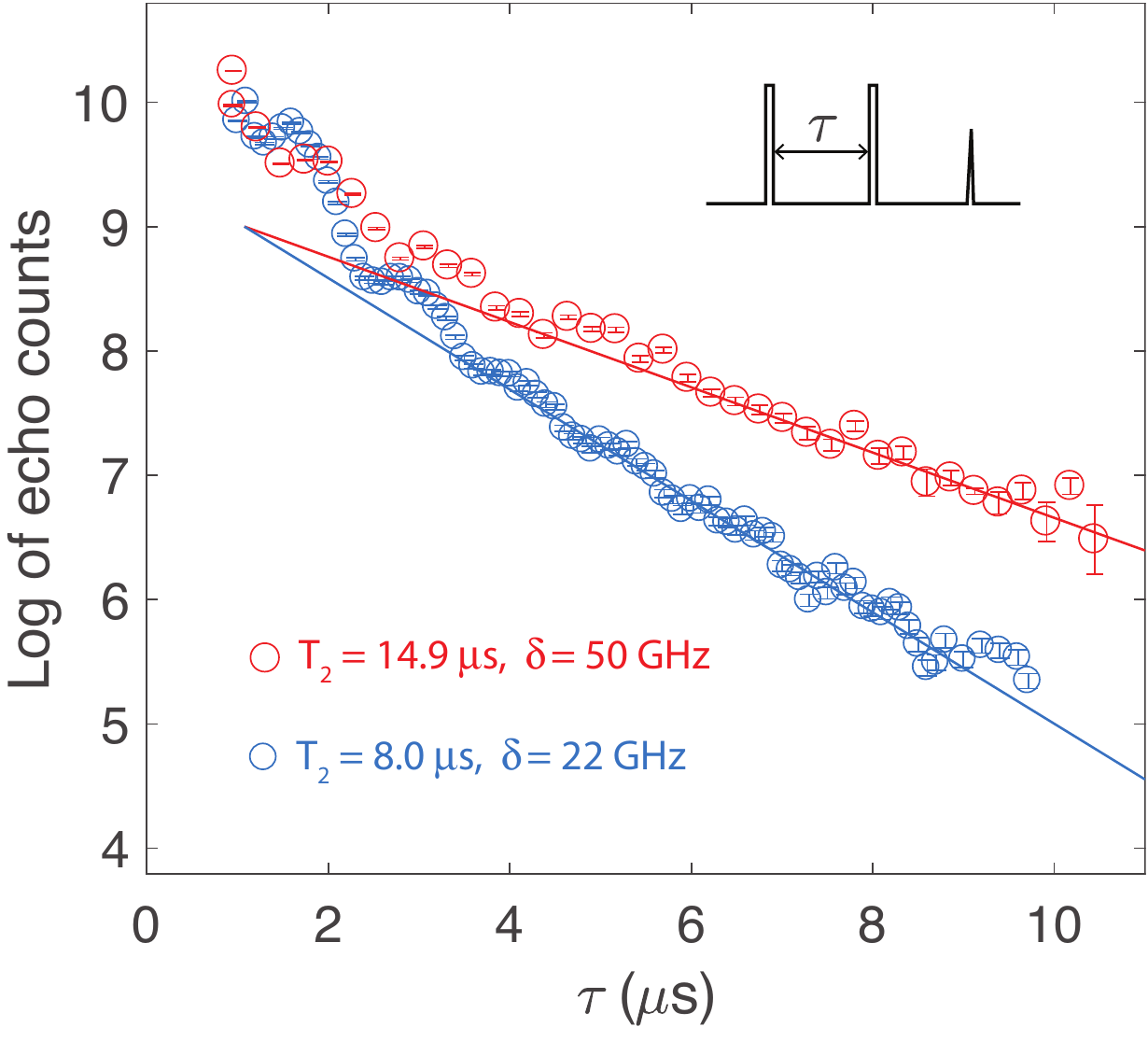}
\caption{\label{figs2} Photon echoes on sub-ensembles of Nd ions coupled to the cavity at different cavity-ensemble detunings. The oscillations evident in the initial echo intensity decays were due to superhyperfine couplings between Nd and Y spins. T$_2$ was fitted from the linear decay sections ($>$4 $\mu$s). A difference in T$_2$ at varying detunings reflects the change of radiative decay rates (i.e.T$_1$) under different Purcell enhancement factors.}
\end{figure}

\section{Modelling superhyperfine couplings}
The superhyperfine interaction is between an electronic spin (Nd in this case) and a neighbouring ligand nuclear spin (yttrium or vanadium). This interaction has been studied in the literature and was typically orders of magnitude weaker than the rare-earth electronic Zeeman interactions (the gyromagnetic ratio of Nd$^{3+}$ is 3.9-33 GHz/T). As a result, the associated Hamiltonian can be treated as a perturbation to the electronic Zeeman coupling of Nd$^{3+}$ under external applied magnetic field \cite{Car}.
\begin{equation}
H' = -{\bf \mu}_{\rm Y}\cdot\left( {\bf B} - \frac{\mu_0}{4\pi}\left[\frac{\langle{\bf {\mu}^{\rm Nd}}\rangle}{r^3} - 3\frac{(\langle {\bf \mu}^{\rm Nd}\rangle\cdot {\bf r})\cdot {\bf r}}{r^5}\right)\right)
\end{equation}
where ${\bf \mu}_{\rm Y}$ is the Y$^{3+}$ nuclear spin magnetic moment, ${\bf B}$ the externally applied magnetic field, ${\bf \mu}^{\rm Nd}$ the Nd$^{3+}$ electronic spin and ${\bf r}$ the vector connecting two spins. The superhyperfine coupling at zero field - the terms in Eq.~(7) that are not dependent on $\bf B$ -  causes each of the Zeeman levels (both ground and excited) to split into a nuclear doublet. Such splitting can be readily calculated from the electronic magnetic dipole moment from the known anisotropic g-factors of Nd. As the applied field increases, eventually the term $-{\bf \mu}_{\rm Y} \cdot {\bf B}$ dominates over other terms and the splitting becomes approximately linear with $\bf B$. Thus, in high fields, the superhyperfine splitting strongly depends on the gyromagnetic ratios of specific ligand nuclear spins. For yttrium it is 2.1 MHz/T. For vanadium it is 11.2 MHz/T. At B = 390 mT along a-axis, the expected superhyperfine splitting of each of the Zeeman branches can be estimated based on atomic coordinates of the yttrium or vanadium ions surrounding the Nd center.

For Nd-Y coupling, there are 4 nearest neighbour Y ions at equal distance of 3.9$\rm \AA$ from each Nd ion. The zero-field superhyperfine splitting is calculated to be $\sim$80 kHz and $\sim$30 kHz for the optical ground and excited levels, respectively. With an applied field of 390 mT along a-axis, the total splitting from Eq.~(7) gives $\Delta_g \sim$740 kHz and $\Delta_e\sim$790 kHz for ground and excited levels.

For Nd-V coupling, the nearest distance between them is 3.14 $\rm \AA$. Since vanadium has 7/2 spins, there are a total of 8 superhyperfine sublevels. The splittings between those levels at 390 mT field ranges from $\sim$4 -30 MHz. Given that the optical excitation pulse only has a bandwidth of 2 MHz, it is unlikely that multiple Nd-V superhyperfine sublevels were excited.

The envelope modulation in Fig.~3(b) can be modelled after derivations in \cite{Mims}, which  include beatings at frequencies $\Delta_e$, $\Delta_g$, ($\Delta_e - \Delta_g$) and ($\Delta_e + \Delta_g$), with their relative strengths dependent on the degree of spin mixing. Due to limited sampling and likely over-simplification of the model, the experimental data (black) in Fig. 3(b) cannot be fitted well with a known analytical form. Instead, an empirical fit (red) was used to identify the dominant beat frequency. The best fit gives 740 kHz, which is in general agreement with the $\Delta_e$, $\Delta_g$ calculated above. We therefore infer that the observed beating was likely originated from the Nd-Y superhyperfine interaction but unlikely from Nd-V couplings. The latter is expected to give a splitting $>$4 MHz. The Nd-V superhyperfine coupling was indeed reported in \cite{deRiedmatten} in the same Nd$^{3+}$:YVO$_4$ material to be about 5 MHz at 0.3 T field.

\subsection{OPTICAL DEPHASING IN Nd$^{3+}$:YVO$_4$}
Possible contributions to the optical dephasing $\gamma^*$=9.7 kHz include superhyperfine coupling between Nd spins and yttrium/vanadium nuclear spins, the Nd spin flip-flops, direct phonon couplings, and other higher order processes. Here we discuss contributions from two potentially dominant mechanisms.

\noindent {\bf Superhyperfine interaction} The experimental condition in the current work closely reassembles that in \cite{Sun} in which optical T$_2$ for a 10 ppm doped Nd$^{3+}$:YVO$_4$ sample was measured at varying magnetic field applied along the a-axis of the crystal. It was found in \cite{Sun} that with a field greater than 1.5 T, the T$_2$ of 27 $\mu$s became limited by the Nd-Vanadium superhyperfine interaction. The corresponding dephasing rate could be calculated from 1/($\pi$T$_2$) - 1/(2$\pi$T$_1$) = 10.0 kHz, which was very close to the dephasing rate measured here. We thus expect that the superhyperfine interactions contribute substantially to the measured $\gamma^*$. Based on the spin interaction models in \cite{Bottger06}, we could numerically estimate the broadenings due to Nd-Y and Nd-V interactions to be 34 and 14 kHz, respectively; these values are in order-of-magnitude agreement with the measurement.

\noindent {\bf Nd spin flip-flops} Dephasing due to the Nd spin flip-flops is a function of the Nd doping concentration and temperature. To better understand this process, we measured optical T$_2$ times in both the 50 ppm doped (the same crystal on which the devices were fabricated) and a nominally undoped YVO$_4$ crystal. From the absorption spectroscopy and secondary ion mass spectroscopy (SIMS), we estimated the doping concentration of Nd to be $\approx$0.2 ppm in the undoped YVO$_4$. Therefore, the dephasing owing to spin flip-flops is expected to be relatively small in that sample. Both crystals were soldered to a common sample holder. With the same magnetic field configuration as in the main text, the ground level splitting was $\mu_Bg_{\perp}{\bf B}$=12.88 GHz where $g_{\perp}$=2.36 is the ground state g-factor \cite{Hastings, AfzeliusYVO}, and ${\bf B}$ = 0.39 T. We then used the ratio between the absorptions of two Zeeman transitions to calibrate the crystal temperature. When both crystals were at $\sim$500 mK, we measured a T$_2^{\rm doped}$ = 25.4 $\mu$s and T$_2^{\rm undoped}$ = 27.0 $\mu$s in 50 ppm doped and undoped YVO$_4$ crystals, respectively. The difference in linewidths, which amounts to $<$1 kHz,  serves as an upper bound on dephasing due to Nd spin flip-flops at 500 mK. 

Using the models put forth by \cite{Bottger06, Bottger09}, the optical dephasing due to Nd-Nd spin flip-flops can be estimated from a Lorentzian spectral diffusion model as $1/\pi T_M$ \cite{Bottger09} where
\begin{equation}
T_M = \frac{2\Gamma_0}{\Gamma_{\rm SD}R}\left(-1+\sqrt{1+\frac{\Gamma_{\rm SD}R}{\pi\Gamma^2_0}}\right),
\end{equation}
where $\Gamma_{\rm SD}$ is  the Nd magnetic dipolar interaction,
\begin{equation}
\Gamma_{\rm {SD}} = \frac{\pi\mu_0|g_g-g_e|g_g\mu^2_B n_{\rm {Nd}}}{9\sqrt{3}\hbar} sech^2(\frac{g_g\mu_B B}{2kT})
\end{equation}
\noindent and $R=1/T_1^{spin}$. Using $n_{\rm {Nd}} = 6.3\times10^{23}$ m$^{-3}$ as the Nd concentration at 50 ppm doping, and Nd spin T$_1^{spin}$=98 ms measured from spectral holeburning at 500 mK, we predict a corresponding optical dephasing to be 30 Hz \cite{Bottger09}. This implies that our measurement is dominated by other interactions, such as the superhyperfine interaction.

\end{document}